\documentclass[a4paper,12pt]{article}

\usepackage[utf8]{inputenc}
\usepackage[T2A]{fontenc}
\usepackage[russian, english]{babel}
\usepackage{amsmath,amssymb}
\usepackage{braket}
\usepackage{array}
\usepackage[unicode,linktocpage=true,plainpages=false,pdfpagelabels=false]{hyperref}

\newcommand{\dd}{\partial}
\newcommand{\de}{\delta}
\newcommand{\al}{\alpha}
\newcommand{\be}{\beta}
\newcommand{\ka}{\varkappa}
\newcommand{\la}{\lambda}
\newcommand{\pp}{\varphi}
\newcommand{\ep}{\varepsilon}
\newcommand{\ga}{\gamma}

\newcommand{\si}{\sigma}

\newcommand{\gMim}{\bar{g}}

\newcommand{\D}{\nabla}

\newcommand{\ls}{\left(}
\newcommand{\rs}{\right)}

\newcommand{\disn}[2]{$$\displaylines{\refstepcounter{equation}%
            \label{#1}\hskip 1em minus 1em #2\hfilneg}$$}
\newcommand{\nom}{\hfil\hskip 1em minus 1em (\theequation)}

\newcommand{\ns}{\hfill\cr\hfill}

\usepackage{amssymb}

\begin{document}

\title{Dual models for p-form mimetic gravity and their connection to perfect fluids consisting of (p+1)-branes
}

\author{R.~V.~Ilin\thanks{E-mail: rireilin@gmail.com},
S.~A.~Paston\thanks{E-mail: pastonsergey@gmail.com}\\
{\it Saint Petersburg State University, Saint Petersburg, Russia}
}
\date{\vskip 15mm}
\maketitle

\begin{abstract}
We propose an approach that allows one to reformulate $n$-dimensional $p$-form mimetic gravity (including usual mimetic gravity as particular case $p = 0$) as nonlinear $(n-p-1)$-form electrodynamics via electric-magnetic duality. The resulting dual Lagrangian density is just the square root of the ordinary quadratic Lagrangian density of $(n-p-1)$-form electrodynamics. By applying field transformation in the action, we show that for the arbitrary $p$ this dual theory transforms into the $(p+1)$-brane fluid: the model of the stack of the parallel $(p+1)$-dimensional branes foliating physical spacetime. As the structure of the field transformations depends on $p$, the sets of solutions in these models are related differently. We prove, that for $p = 0$ and $p = n-2$ dual mimetic models describe usual particle fluid with the potential flow and to the $(n-1)$-brane fluid respectively. For other values of $p$ not all mimetic solutions behave like that, in general, so we restrict ourselves only to the case $n = 4$, $p = 1$. In this case, we show, that mimetic formulation is dual to the well-known Nielsen-Olesen theory of "dual strings" and discuss the criterion indicating whether its solutions behave like string fluid. The cosmological solutions for these models in are also discussed.
\end{abstract}

\section{Introduction}
\sloppy
\label{Introduction}

One of the most prominent and simple models explaining the dark matter phenomenon that was proposed in the last decade is mimetic gravity. The interest in this model has been increasing over time (see \cite{mimetic-review17, OdintsovNojiriMimetic2014} and references therein), and now it is still alive even in the light of the new observations \cite{MimeticAliveAndWell2018}. The original idea \cite{mukhanov} was to isolate the conformal mode of the metric with help of singular disformal transformation in the Einstein-Hilbert action:
\begin{equation}\label{sp1}
    g_{\mu\nu} = \bar{g}_{\mu\nu}\bar{g}^{\al\be}\dd_\al\pp\dd_\be\pp,
\end{equation}
where $\bar{g}_{\mu\nu}$ is auxiliary metric, and $\pp$ is new scalar field. The degeneracy of this transformation comes from the invariance of this expression under the local Weyl transformations of the auxiliary metric $\gMim_{\mu\nu}$. This property is crucial: in \cite{Golovnev201439} it was discussed, that this degeneracy and presence of the derivatives of $\pp$ constraint the variation of the physical metric $\de g_{\mu\nu}$. Therefore, there are fewer restrictions on the physical metric $g_{\mu\nu}$ imposed by the equations of motion (EOM) compared to General Relativity. This is exactly the source of the new degrees of freedom, that effectively can be used to describe dark matter. In the same work, mimetic gravity was formulated in an alternative way as usual General Relativity with the pressureless potentially moving dust. It allows one to successfully apply this model and its numerous modifications to fit the observational data, see, for example, \cite{MimeticMONDLikeVagnozzi, MimeticStaticSpherSymm2015, NojiriMimeticCompletion2016, OdintsReconstructionMimetic2018}. However, unlike the model with the transformation \eqref{sp1}, in most of these modifications the dark matter does not have clear physical interpretation.

In the present paper, we step forward in that direction and study $p$-form mimetic gravity in the $n$-dimensional spacetime. It was first considered in the work \cite{F2AbelianMimeticGorjiMukhoyama} for the case $n = 4$. In the section \ref{DualAlgorithm} we show, that by using the $p$-form analogue of the electric-magnetic duality one can reformulate this model as the "square root" of $(n-p-1)$-form electrodynamics. This dual formulation does not contain any Lagrange multipliers, which nearly all mimetic models have (see, for example, \cite{Golovnev201439,statja48}). In the section \ref{razb} we show, that by using certain differential transformations in action, the dual model transforms into the perfect fluid of $(p+1)$-dimensional branes, that can be described within the splitting theory approach \cite{statja25} (see also \cite{BorlafBraneFluid}). This transformation relates the sets of solutions of EOM in both theories, just like the transformation \eqref{sp1} in the original mimetic gravity. In section \ref{cases} we study this relation in detail for the most interesting particular cases, and finally, in the last section 5 we discuss the cosmological applications for the analysed theories for the case $n = 4$.

\section{Dual action for p-form mimetic gravity}\label{DualAlgorithm}
To formulate $p$-form mimetic gravity, instead of transformation \eqref{sp1} we consider the following transformation in the $n$-dimensional Einstein-Hilbert action $S_{EH}$:
\begin{equation}\label{sp2}
g_{\mu\nu} = \bar{g}_{\mu\nu}\ls
\be
F_{\mu_1\dots\mu_{p+1}}F_{\nu_1\dots\nu_{p+1}}
\bar{g}^{\mu_1\nu_1}\dots\bar{g}^{\mu_{p+1}\nu_{p+1}}\rs^{\frac{1}{p+1}},
\end{equation}
where $\mu,\nu,\ldots=0,\ldots,n-1$.
Here $F_{\mu_1\dots\mu_{p+1}}$ is field strength $(p+1)$-form with $p$-form potential $A_{\mu_1\dots\mu_p}$ (i.e. $F = dA$), and $\be = \pm1$. Like the transformation \eqref{sp1} this formula is also Weyl-invariant. We will everywhere assume, that the quantity in the brackets in \eqref{sp2} is non-negative, which constrains the potential $A$. If one takes $p=0$, $\be = 1$ it is obvious, that the transformation \eqref{sp2} coincides with \eqref{sp1} (we are working with the signature of metric $+-\dots-$).

It can be easily shown, that this theory can be reformulated as GR with the additional mimetic matter as it was done for the transformation \eqref{sp1} in \cite{Golovnev201439} (see also ~\cite{F2AbelianMimeticGorjiMukhoyama} for $n=4$):
\begin{equation}
S =S_{EH}+S_{mim},\qquad
S_{mim}[g_{\mu\nu}, A]=  \frac{1}{2}  \int d^nx\sqrt{|g|}\la\ls \langle F, F\rangle_g-\be
    \rs.
    \label{OriginalPFormMimeticAction}
\end{equation}
Here we introduced the Lagrange multiplier $\la$ and used the notation for the contraction of the two $(p+1)$-forms:
\disn{spn1}{
\langle F, F\rangle_g=\frac{1}{(p+1)!}\,F_{\mu_1\dots\mu_{p+1}}F_{\nu_1\dots\nu_{p+1}}
{g}^{\mu_1\nu_1}\dots{g}^{\mu_{p+1}\nu_{p+1}}.
\nom}
By varying this action with respect to $\la$ and metric $g_{\mu\nu}$, one obtaines the following equations of motion:
\begin{align}
    & \langle F, F\rangle_g=\be, \label{sp4}\\
    & G_{\mu\nu} = \ka T_{\mu\nu},\qquad T^{\mu\nu} = \frac{1}{p!}\,\la\,  F^{\mu\mu_1\dots\mu_p}F^{\nu}{}_{\mu_1\dots\mu_p}. \label{EMT0}
\end{align}

In the following, it will be very convenient to use the standard differential form notations and the operations from the exterior algebra associated with them. In these terms $S_{mim}$ can be rewritten
\disn{spn2}{
S_{mim}[g_{\mu\nu}, A]=\frac{1}{2}\int\la\left(\sigma F\wedge \star F-\be\star\! 1\right),
\nom}
where $\si = (-1)^{n-1}$ is the sign of the metric determinant. The last EOM is obtained by varying $S_{mim}$ with respect to $A$:
\disn{sp3}{
d(\la \star\! F)=0.
\nom}

Our next goal is to obtain a new formulation of this theory via certain analog of the electric-magnetic duality. Though one may think, that this duality works similarly to the ordinary $p$-form electrodynamics, we will show, that it is quite different in the mimetic case. We assume that the spacetime topology is trivial in the following. Then, by virtue of the Poincare lemma, the equation \eqref{sp3} is equivalent to the existence of the $q$-form potential $W$ ($q\equiv n-p-2$, we are assuming that $p < n-2$), that is defined below:
\disn{sp5}{
\la \star\! F=dW\equiv B.
\nom}
Now one can rewrite other equations of motion only in terms of $W$. As contraction \eqref{spn1} has the property $\langle \star F, \star F\rangle_g=\si \langle F, F\rangle_g$, the equation \eqref{sp4} can be rewritten as
\disn{sp6}{
\langle B, B\rangle_g=\si\be\la^2.
\nom}
In terms of $B$ the equation \eqref{sp3} is automatically satisfied. However, one must take into account the Bianchi identities for $F$ and use them as the new equations of motion for $B$:
\disn{sp7}{
dF = d\left(\la^{-1}\star B\right)=0.
\nom}
Now we can solve the equations \eqref{sp6} and \eqref{sp7} with respect to $\la$:
\disn{DualGeneralEOM}{
\nabla_\mu\left(\frac{B^{\mu\mu_1\dots\mu_{q}}}{\sqrt{\si\be \langle B, B\rangle_g}}\right) = 0,
\nom}
where $\nabla_\mu$ is covariant derivative with Levi-Civita connection for the physical metric $g_{\mu\nu}$. Lastly, by using \eqref{sp7} and \eqref{sp5} one can easily recast \eqref{EMT0} in terms of $B$ only.

To sum up, the $p$-form mimetic gravity can be reformulated as the theory of $q$-form potential $W$ instead of $A$ with the equations of motion being only \eqref{DualGeneralEOM} and the Einstein's equations \eqref{EMT0} rewritten in terms of $W$. It is not hard to find the corresponding variational principle for such theory by taking the following as $S_{mim}$:
\begin{equation}
S_{mim}[g_{\mu\nu}, W]=-\int d^n x\sqrt{|g|}\sqrt{\si\be \langle B, B\rangle_g}.
\label{BraneCloudAction}
\end{equation}
Indeed, by using the notation \eqref{spn1} the variation of this action with respect to $W$ leads directly to \eqref{DualGeneralEOM}. The expression for EMT can also be verified to coincide with \eqref{EMT0}, where $\la$ and $F$ are expressed in terms of $B$ according to \eqref{sp6} and \eqref{sp5}. In the following, we will call \eqref{BraneCloudAction} a dual formulation of $p$-form mimetic gravity.\par

As it was noted above, the resulting dual theory is quite different from what one may expect from the theory similar to the $p$-form electrodynamics. The key reason for such behavior is the presence of the equation \eqref{sp4}. Indeed, if one omits this equation and considers only the Maxwell-like equation \eqref{sp3}, there exists a natural generalization of the standard duality transformation from electrodynamics. Namely, if one performs $\lambda' =\la^{-1}, F' = \la\star F$, it can be easily seen, that \eqref{sp3} and Bianchi identities in $p$-form and $(n-p-2)$-form cases are transformed into each other. However, this transformation applied for \eqref{sp4} in $p$-form mimetic theory does not lead to the equation \eqref{sp4} in $(n-p-2)$-form theory as can be seen by a simple calculation. Thus, mimetic $p$-form and $(n-p-2)$-form theories are indeed not equivalent, and in the next sections we will further justify this by illustrating, that these cases have different physical content.

It should be noted, that we fixed the total sign as $"-"$, which corresponds to a certain branch of the solution \eqref{sp6} in the original theory. Hence, the action \eqref{BraneCloudAction} describes only "one half" of the original mimetic theory \eqref{OriginalPFormMimeticAction}. This choice is beneficial in ordinary mimetic gravity ($p = 0$) as it does not violate the dominant energy condition for the dark matter (unlike the opposite sign). In the next section, we will show, that this choice is also natural in the light of the connection of this model with the perfect fluid of $(p+1)$-branes.

\section{Connection with brane fluid (cloud) models}\label{razb}
The action \eqref{BraneCloudAction} has some obvious advantages over the initial action \eqref{OriginalPFormMimeticAction}. It does not contain any Lagrange multipliers and is also a negative-definite. But also has much clearer physical meaning. In the current section we will show, that the action \eqref{OriginalPFormMimeticAction} can be related to the theory of non-interacting stack of $(p+1)$-branes that form a foliation of spacetime with the action in the form described in \cite{statja25}. This theory can be understood as perfect "fluid" (or "cloud") consisting of the $(p+1)$-branes (in the sense of works \cite{LetelierStringCloud, StachelStringDust1, GibbonsStringFluid2001, BorlafBraneFluid}).

To derive the action of this theory, we assume that the spacetime $M$ with coordinates $x^\mu$ is foliated by the level set of non-intersecting branes $M_h$ of dimension $\dim M_h = n-m$ with the level functions $h^A$ (hereinafter $A,B,\ldots=1,\ldots,m$). On each brane we define such coordinates $\hat x^i$ ($i,k,\ldots=0,\ldots,n-m-1$), that for the fixed $h^A$ the functions $x^\mu(\hat x^i)$ define smooth embeddings of $M_h$ into $M$. The total action of this system is then the sum of the actions of the separate branes:
\disn{sp11}{
S_{f}[g_{\mu\nu}, x^\mu]=-\int d^m h \int_{M_h} d^{n-m}\hat x\sqrt{|\hat g|},
\nom}
where $\hat g=\det\hat g_{ik}$, and $\hat g_{ik}$ is the induced metric on the brane:
\disn{sp10}{
\hat g_{ik}=\frac{\dd x^\mu}{\dd\hat x^i}g_{\mu\nu}\frac{\dd x^\nu}{\dd\hat x^k}.
\nom}

To relate this action to the \eqref{BraneCloudAction}, one must find a more convenient formulation. Note, that one can use the set $\tilde x^\mu\equiv\{\hat x^i,h^A\}$ as the coordinates on $M$. Therefore, it is possible to perform a diffeomorphism from these special coordinates $\tilde x^\mu$ to the arbitrary ones $x^\mu$. The method of calculation of the corresponding Jacobian can be found in \cite{statja42}, for the current needs it yields:
\disn{sp9}{
\left|\det \frac{\dd \tilde x^\mu}{\dd x^\nu}\right|=\sqrt{\frac{gw}{\hat g}},
\nom}
where $w=\det(w^{AB})$, and $w^{AB}=g^{\mu\nu}(\dd_\mu h^A)(\dd_\nu h^B)$. After applying the diffeomorphism, the action \eqref{sp10} takes the form:
\disn{sp14}{
S_{split}[g_{\mu\nu}, h^A] = -\int_M d^n x\sqrt{|g|}\sqrt{(-1)^{m}w}.
\nom}
It should be stressed, that unlike \eqref{sp11} here $h^A$ are the arguments of the action instead of the embedding functions $x^\mu(\hat{x}^i)$ of the individual branes $M_h$. In the work \cite{statja25} (there this description of the parallel branes was called splitting theory) it was proven, that theories \eqref{sp14} and \eqref{sp11} are equivalent at the level of the equations of motion. The difference between the specified and the present works is, that in \eqref{sp11} $M$ was flat bulk spacetime, and the dynamics of $M_h$ was described by Regge-Teitelboim embedding gravity \cite{regge} with Einstein-Hilbert action.

The determinant of $w^{AB}$ by definition can be written in the form:
\disn{sp15}{
w=\frac{1}{m!}
\ep_{A_1\dots A_m}(\dd_{\mu_1}h^{A_1})\ldots(\dd_{\mu_{m}}h^{A_{m}})
\ep_{B_1\dots B_m}(\dd_{\nu_1}h^{B_1})\ldots(\dd_{\nu_{m}}h^{B_{m}})\times\ns\times
g^{\mu_1\nu_1}\ldots g^{\mu_{m}\nu_{m}},
\nom}
or, by using the notation \eqref{spn1}:
\disn{sp16}{
w=
\langle dU, dU\rangle_g,\qquad
U[h^A]=\frac{1}{m!}\ep_{A_1\dots A_{m}}h^{A_1} dh^{A_2}\wedge\ldots\wedge dh^{A_{m}}.
\nom}
Thus, the action \eqref{sp14} has the same form as dual mimetic $p$-form action \eqref{BraneCloudAction}, if one takes $m=q+1$, $\be=(-1)^{n-q}$ and express the independent potential $W$ as it is prescribed in \eqref{sp16}. This is the previously announced result: the action \eqref{sp14} is related to the \eqref{BraneCloudAction} by virtue of the differential transformation in action. Now it is also clear, that the total minus sign in \eqref{BraneCloudAction} is natural there as it provides the link to the brane fluid model.

Before moving forward, we want to draw attention to the simple fact: if $m>1$ the transformation $W\rightarrow U[h^A]$ from \eqref{sp16} contains derivatives of $h^A$. As was discussed in \cite{statja60, Golovnev201439, DeserJakiwClebsch2001}, their presence can alter the set of solutions of the equations of motion compared to the non-transformed theory. Therefore, it is correct to assume, that not all dual mimetic solutions in theory \eqref{BraneCloudAction} are also brane fluid solutions for \eqref{sp11}. In the rest of the article, we study this relation between the solutions of two theories. It will be shown, that for the specific values of $p$ {\it all} $p$-form mimetic gravity solutions behave like $(p+1)$-brane fluid.

\section{When does $p$-form mimetic gravity behave like brane fluid?}\label{cases}
Firstly, consider the original mimetic gravity, which corresponds to $p = 0$. It can be shown, that in the case $\be = -1$ the dominant energy condition is violated, so we will only consider $\be = 1$. As it was stated above, it is well-known, that such theory describes potentially moving pressureless dust. It is instructive to show it using the results of the previous section.

In this case $q = n-2$ and hence $\dim M_h = 1$, i.e. branes are just the particle worldlines. For $\be=1$ the action \eqref{sp11} and therefore \eqref{sp14} then describe perfect fluid without pressure. The corresponding dual mimetic action is related to it with the change \eqref{sp16}. To understand, how the solutions of EOMs in both theories are related, it is instructive to write the equations of motion through $4$-velocity, that depends on the potential form $W$:
\begin{equation}
    u[W] \equiv \frac{1}{\sqrt{\sigma\langle dW,dW\rangle_g}}\star dW.
    \label{uDef}
\end{equation}
For theory \eqref{BraneCloudAction} EOMs are the following:
\begin{equation}
    du[W] = 0,
    \label{p0MimeticEq}
\end{equation}
while for the action \eqref{sp14} they can be written as:
\begin{equation}
    dh^{A_1}\wedge\dots\wedge dh^{A_{n-2}}\wedge du[U] = 0,
    \label{ParticleSplittingEquation}
\end{equation}
where in these formulae $U$ is defined in \eqref{sp16}, and $W$ was previously introduced in \eqref{sp5}. To proceed further one can use the generalization of well-known Clebsch representation \cite{Clebsch1859} for the arbitrary $(n-2)$-form \cite{RundDiffgeomTopicsClebsch} in \eqref{p0MimeticEq}:
\disn{sp17}{
W=d\al+ \frac{1}{(n-1)!}\ep_{BA_1\dots A_{n-2}}k^{B} dk^{A_1}\wedge\ldots\wedge dk^{A_{n-2}},
\nom}
where $\al$ and $k^A$ are $(n-3)$-form and $0$-forms respectively. The first term in this representation does not contribute to \eqref{p0MimeticEq}, and the second form is simply $U[k^A]$. Taking this into account, it is now obvious, that all dual mimetic solutions of the equations \eqref{p0MimeticEq} are also solutions of the system \eqref{ParticleSplittingEquation} and hence they behave like pressureless dust. The equations \eqref{p0MimeticEq} also impose additional conditions on $h^A$ compared to \eqref{ParticleSplittingEquation}. Indeed, from \eqref{uDef} it follows, that $u = d\pp$. It is precisely the requirement of dust flow being potential, which was mentioned above.

It is worth noting, that the action \eqref{BraneCloudAction} for $p = 0$ can be easily connected with alternative formulations of mimetic gravity. As an example, one can rewrite \eqref{BraneCloudAction} in terms of auxiliary current $j^\mu$:
\disn{sp12}{
S_{mim}[g_{\mu\nu}, W]=-\int d^n x\sqrt{|g|}\sqrt{j^\mu j^\nu g_{\mu\nu}},\qquad j\equiv\star B.
\nom}
If the spacetime topology is trivial, the condition $B = dW$ (see \eqref{sp8}) is equivalent to the conservation law $d\star j = \nabla_\mu j^\mu = 0$. This leads to the new action, where $j^\mu$ is independent field, and its conservation enters the action with Lagrange multiplier $\tau$:
\disn{sp13}{
S_{mim}[g_{\mu\nu}, j^\mu,\tau]=-\int d^n x\sqrt{|g|}\ls\sqrt{j^\mu j^\nu g_{\mu\nu}}+\tau \nabla_\mu j^\mu\rs.
\nom}
This is the action for mimetic gravity $p = 0$, $n = 4$, that was discussed in \cite{statja48}.

Next we consider another edge case $p = n-2$, which maximum allowed value for $p$. Then $q = 0$, and the potential $W$ is scalar. For simplicity, assume that $\be = (-1)^n$. In this case \eqref{BraneCloudAction} greatly simplifies:
\disn{sp8}{
S_{mim}[g_{\mu\nu}, W]=-\int d^n x\sqrt{|g|}\sqrt{-(\dd_\mu W)(\dd_\nu W)g^{\mu\nu}}.
\nom}
In this case, $m = 1$, and hence the transformation \eqref{sp16} doesn't have derivatives of $h$, moreover, it is simply $W = h$. Thus, the dual mimetic theory \eqref{sp8} is exactly equivalent to the perfect fluid of $(n-1)$-branes, described by \eqref{sp14} for $p = n-2$, $\be = (-1)^n$.

 In the work \cite{F2AbelianMimeticGorjiMukhoyama} the particular case $p = 2, n = 4$ of theory \eqref{sp8} was considered, and it was stated, that this theory is equivalent to the usual mimetic gravity $p = 0$. The equivalence between $p = 2$ and $p = 0$ does exist in the ordinary $p$-form, but for {\it mimetic} $p$-form theory with $\be = (-1)^n$ it is not true, as we discussed after \eqref{sp8}. However, we didn't inspect the case $\be = (-1)^{n-1}$, so one might suggest, that the equivalence with case $p = 0$ may still take place for this choice of $\be$. In this model positivity of the expression under the square root in \eqref{sp8} leads to $M_h$ defined in \ref{razb} being spacelike, therefore these surfaces are not branes. Instead, one may interpret lines, that are normal to $M_h$ as worldlines of the particles. In this case, the action \eqref{BraneCloudAction} will differ from \eqref{sp8} only by the sign under the square root (and possibly, by the total sign). By introducing $4$-velocity similarly to the case $p = 0$ (see \eqref{uDef}), one can write the equations of motion in the simple form:
\disn{sp8a1}{
\D_\mu u^\mu=0,\qquad u_\mu \equiv \frac{1}{\sqrt{g^{\al\be}\dd_\al W\dd_\be W}}\dd_\mu W.
\nom}
By using \eqref{sp8} it is easy to show, that the corresponding energy-momentum (EMT) tensor is EMT for (exotic) perfect fluid with $\rho = 0$ and non-zero pressure, which obviously violates dominant energy conditions. The nature of such EMT structure is quite simple. Though energy density is zero, one can always define particle density $\nu$ at a given moment of time, which satisfies the continuity equation $\D_\mu(\nu u^\mu)=0$. From \eqref{sp8a1} it then follows, that $u^\mu\dd_\mu \nu= 0$, so the particle density $\nu$ doesn't change along worldlines. Thus, the case $p = n-2$, $\be = (-1)^{n-1}$ describes incompressible fluid with non-geodesic motion, which drastically differs from the case $p = 0$, and hence there is no equivalence between the two theories regardless of the sign of $\be$.

For other values of $p$ the situation is more complicated. The simplest model of this kind is $p = 1$, $n = 4$, which cosmological applications (with some reasonable moditifications, see section \ref{CosmSection}) were studied in \cite{F2AbelianMimeticGorjiMukhoyama}. We will restrict our attention only to $\be = -1$. For such theory $q = 1$, so the action \eqref{BraneCloudAction} takes the form:
\begin{equation}
    S_{mim}[g_{\mu\nu}, W_\mu] = -\int d^4x\sqrt{-g}\sqrt{B_{\mu\nu}B^{\mu\nu}}.
    \label{n4p1NielsenOlesenAction}
\end{equation}
This action was already discussed in the literature and is known as the "dual strings model" proposed by Nielsen and Olesen \cite{NielsenOlesenFieldStringTheory1973}. In this case $m = 2$, i.e. the action \eqref{sp11} corresponds to the Nambu strings fluid. The corresponding change of variables \eqref{sp16} has the following explicit form:
\disn{sp18}{
U=\frac{1}{2}\ep_{AB}h^{A} d h^{B}.
\nom}
This change is very different from the cases discussed above. Namely, consider dual mimetic theory with $p = 0$. It is obvious, that if for any $\al$ one performs the transformation of the form $W\rightarrow U'[\al, h^A] \equiv d\al + U[h^A]$ in \eqref{BraneCloudAction}, the resulting action will not depend on $\al$. Therefore one may consider $U'$ as a transformation that links \eqref{sp11} and \eqref{sp14} instead of $U$. On the other hand, $U'[\al, h^A]$ is just the Clebsch representation (see \eqref{sp17}) of the {\it arbitrary} $(n-2)$-form, and thus any $(n-2)$-form potential $W$ can be written as $U'[\al, h^A]$ for some $\al$ and $h^A$. The same line goes for the case $p = n-2$, where $U'$ is simply trivial (see text after \eqref{sp8}). However, for $p = 1$, $n = 4$ it is clear from \eqref{sp18}, that in this case not for every potential $W$ the appropriate $\al$ and $h^A$ can be found. From the perspective of the equations of motion, it means, that not all solutions in \eqref{n4p1NielsenOlesenAction} behave like string fluid. In addition, those who do are not moving arbitrarily: as in the case $p = 0$ their movement will obey potentiality conditions.

There is another simple criterion, that allows one to determine if specific solution behave like string fluid or not. To derive it one should look at the field strength $\tilde{B}$, which corresponds to the potential $U$, defined in \eqref{sp18}:
\disn{sp19}{
\tilde{B}=\frac{1}{2}\ep_{AB}dh^{A} \wedge d h^{B}.
\nom}
This field strength does not only satisfy Bianchi identity but also the relation $\tilde{B}\wedge \tilde{B}=0$ (it was discussed in \cite{GibbonsStringFluid2001}). It means, that at every point $\tilde{B}$ is proportional to the volume form of some surface (which is not necessarily the same at each point). In general, field strength $B$ in dual mimetic model \eqref{n4p1NielsenOlesenAction} does not have this property. By using the "electric" ${\bf E}$ and "magnetic" ${\bf H}$ components of $B$, this condition can be written as ${\bf E}\cdot {\bf H} = 0$. In particular, it is satisfied when either ${\bf E}$ or ${\bf H}$ are zero.

\section{Cosmology for $n = 4$}\label{CosmSection}
Lastly, we want to briefly discuss the cosmological applications of the presented models. Once again, we will only consider the case $n = 4$, so the only interesting models are $p = 0,1,2$. The FLRW solutions for $p = 0$ are well-known and are just models of the universe filled with pressureless dust, so $p = 0$. For the other two cases the situation is more complicated. In case $p = 1$ it can be shown, that any vector field satisfying \eqref{DualGeneralEOM} breaks rotational symmetry and, moreover, none of these satisfy the Einstein's equations. There are several possible workarounds, but we will stop at the one used in \cite{F2AbelianMimeticGorjiMukhoyama}, where the case $p = 1, \be = -1$ was considered. Namely, the idea is to modify the theory by introducing additional vector fields to $A_\mu$ and turning them to the global $\text{SO}(3)$-triplet $A_\mu^a$, where $a = 1,2,3$ (similarly to so-called cosmic triad \cite{GolovnevMukhanovTriad_2008}). The action of this model is required to be $\text{SO}(3)$ rotations of this triplet and is just the simplest generalization of the action \eqref{n4p1NielsenOlesenAction} of this kind:
\begin{equation}
    S_{mim}[g_{\mu\nu}, W_\mu] = -\int d^4x\sqrt{-g}\sqrt{B^a_{\mu\nu}B^{a\mu\nu}},
\end{equation}
where $B^a_{\mu\nu}\equiv \dd_\mu W_\nu^a - \dd_\nu W_\mu^a$.
As it was shown in \cite{F2AbelianMimeticGorjiMukhoyama} for the original theory \eqref{OriginalPFormMimeticAction} for the spatially flat Friedman geometry the solution exists, and the mimetic matter behave like the perfect fluid with the equation of state $P = (-1/3)\rho$ (here $P$ denotes pressure) while the vector field takes the form $A^a_\mu = \zeta t^2\de^a_\mu $ and $\lambda = \gamma
 t^{-2}$, where $\xi, \ga$ are constants. The great advantage of such modification is that it does not break the dual action construction, that was described in the section \ref{DualAlgorithm}. Indeed, the action for the dual theory will have exactly the same form \eqref{BraneCloudAction} with the only difference being the triplet index of the field strength $B^a_{\mu\nu}$. The dual FLRW-solutions are $B^a_{0i} = 0,\; B^a_{ij} = \text{const}$.\par
 
 The analysis of FLRW-solutions for $p = 2$ is very similar. Unlike the previous case, it is convenient to seek the solutions using dual frame \eqref{sp8}. Recall, that in this case the dual potential $W$ is just scalar field. Although, the construction of FLRW-solutions in the case $\be = -1$ is straightforward due to the vector $\dd_\mu W$ being time-like, the mimetic matter in this theory will violate dominant energy condition as we have seen in the previous section (see discussion after \eqref{sp8}). Therefore we will only focus on the case $\be = 1$, so the gradient $\dd_\mu W$ is assumed to be space-like. It is obvious, that in this case $\dd_\mu W$ violates the spatial isotropy, and one can easily prove, that there are no FLRW-solutions in the standard version of this theory. Just like in case $p = 1$, one may resolve the issue by generalizing scalar $W$ to the $\text{SO}(3)$-triplet $W^a$ and requiring the action to be invariant under the global $\text{SO}(3)$-transformations:
 \begin{equation}
     S_{mim}[g_{\mu\nu}, W] = -\int d^4x\sqrt{-g}\sqrt{-g^{\mu\nu}\dd_\mu W^a\dd_\nu W^a}.
\label{p2GenAction}
 \end{equation}
The corresponding energy-momentum tensor has the following structure:
\begin{equation}
T_{\mu\nu}= \sqrt{-g^{\al\be}\dd_\be W^b\dd_\al W^b}\left[g_{\mu\nu} - \frac{\dd_\mu W^a\dd_\nu W^a}{g^{\al\be}\dd_\be W^b\dd_\al W^b}\right],
\label{p2EMT}
\end{equation}
while the the equations of motion, that can be obtained by varying \eqref{p2GenAction} with respect to $W^a$ take the form:
\begin{equation}
\dd_\mu\left(\sqrt{-g}\frac{g^{\mu\ga}\dd_\ga W^a}{\sqrt{-g^{\al\be}\dd_\be W^b\dd_\al W^b}}\right) = 0.
\label{p2SO3EOM}
\end{equation}
One can verify the existence of FLRW solutions in this theory by using the ansatz $W^a = C\de^a_\mu x^\mu$, where $C = \text{const}$, and $x^\mu$ are space-time coordinates, and choosing the Friedmann universe to be spatially flat. In this case $\dd_\mu W^a = C\de^a_\mu$, and therefore this expression for $W^a$ satisfies the equations of motion  \eqref{p2SO3EOM}. It can be easily seen then, that if one substitue this ansatz in \eqref{p2EMT}, the resulting expression will take the form of the energy-momentum tensor of ideal fluid with equation of state $P = (-2/3)\rho$.

\section*{Acknowledgements}
The work of the authors is supported by RFBR grant № 20-01-00081.

\section*{Conflict of Interest}
The authors declare no conflict of interest.

\end{document}